%% file: main.tex
\documentclass[letterpaper,twocolumn,10pt]{article}
\usepackage{usenix2019_v3}

\usepackage{tikz}
\usepackage{amsmath}

\usepackage{filecontents}

\usepackage{xspace}
\usepackage{subcaption}
\usepackage{algorithmic}
\usepackage[linesnumbered,ruled,vlined]{algorithm2e}
\usepackage{enumitem} 
\usepackage{graphicx}
\usepackage{booktabs}
\usepackage[dvipsnames]{xcolor}
\usepackage{authblk}
\usepackage{amsmath}

\usepackage{float}
\usepackage{multirow}
\usepackage{amssymb} 
\usepackage{wrapfig}

\newcommand{\proj}{Mesh-Attention\xspace}
\newcommand{\projabbr}{Mesh\xspace}
\newcommand{\attnmaxspeedupring}{32.27$\times$\xspace}
\newcommand{\attnmaxspeedupusp}{4.36$\times$\xspace}
\newcommand{\attnmaxspeedupstartrail}{4.59$\times$\xspace}

\newcommand{\attnavgspeedupring}{17.00$\times$\xspace}
\newcommand{\attnavgspeedupusp}{2.62$\times$\xspace}
\newcommand{\attnavgspeedupstartrail}{2.94$\times$\xspace}

\newcommand{\rev}[1]{\textcolor{black}{#1}}
\newcommand{\red}[1]{\textcolor{black}{#1}}

\begin{document}

\date{}

\title{\Large \bf \proj: A New Communication-Efficient Distributed Attention with Improved Data Locality}


\makeatletter
\renewcommand\AB@affilsepx{, \protect\Affilfont}
\makeatother



\author{
{\rm Sirui Chen\textsuperscript{‡*} }
{\rm Jingji Chen\textsuperscript{§*} }
{\rm Siqi Zhu\textsuperscript{¶††} }
{\rm Ziheng Jiang\textsuperscript{§} }
{\rm Yanghua Peng\textsuperscript{§**} }
{\rm Xuehai Qian\textsuperscript{†**} }
\\
{\rm \textsuperscript{†}Tsinghua University }
{\rm \textsuperscript{‡}Purdue University }
{\rm \textsuperscript{¶}University of Illinois Urbana-Champaign }
{\rm \textsuperscript{§}ByteDance Seed }
} 

\maketitle

\def\thefootnote{*}\footnotetext{Equal contribution.}\def\thefootnote{\arabic{footnote}}
\def\thefootnote{††}\footnotetext{Work done during internship at ByteDance Seed.}\def\thefootnote{\arabic{footnote}}
\def\thefootnote{**}\footnotetext{Corresponding authors.}\def\thefootnote{\arabic{footnote}}

\begin{abstract}

Distributed attention is essential for scaling large language models (LLMs) to long contexts, yet existing methods either have limited parallelism or incur high communication costs. Ulysses uses efficient all-to-all communication but cannot scale beyond the number of attention heads, whereas Ring-Attention removes this limit at the cost of high per-GPU communication. In our analytical Q-KV interaction matrix model, Ring-Attention assigns each GPU an entire row, preserving Q locality while sacrificing KV locality. Each GPU therefore receives nearly all KV partitions, and its communication grows linearly with sequence length. We present \proj, which instead assigns each GPU a 2D tile to balance Q and KV locality, so that it collects only subsets of Q and KV partitions. This gives \proj asymptotically lower communication complexity than Ring-Attention without limiting parallelism. \proj further uses KV Partition Rotation (KVR), greedy scheduling, and topology-aware GPU mapping to balance traffic, overlap communication with computation, and reduce traffic over low-bandwidth links. 
\red{Across experiments on up to 256 GPUs and 1M-token sequences, \proj achieves average speedups of \attnavgspeedupring, \attnavgspeedupusp and \attnavgspeedupstartrail  (up to \attnmaxspeedupring, \attnmaxspeedupusp and \attnmaxspeedupstartrail) over Ring-Attention, USP (Ulysses degree 8) and StarTrail, respectively.} It maintains this performance advantage as GPU count and sequence length increase while substantially reducing communication overhead at scale.

\end{abstract}

\input{intro}

\input{background}
\input{method}

\input{experiment}
\input{conclusion}

\bibliographystyle{plain}
\bibliography{refs}

\end{document}

%% file: intro.tex
\section{Introduction}

Large language models (LLMs)~\cite{zhao2025surveylargelanguagemodels} 
show impressive capabilities in completing various
real-world tasks such as AI agents~\cite{liu2025agentbenchevaluatingllmsagents}, document summarization~\cite{zhang2025comprehensivesurveyprocessorientedautomatic}, virtual assistants~\cite{openai_gpt4techreport}, video understanding~\cite{zhang2023videollamainstructiontunedaudiovisuallanguage,cheng2024videollama2advancingspatialtemporal}, and code
completion~\cite{10.1145/3491101.3519665}. 
For LLMs, it is important to scale to larger context window 
(i.e., the maximum number of tokens that can be 
processed by the model) so that a greater amount of information
(e.g., longer documents or videos) can be leveraged 
when handling tasks.
For this reason, the size of context window is a crucial
metric when comparing different LLMs, and supporting longer
context window has become an appealing selling point. 
For example, \rev{Gemini 3.1 Pro}~\cite{gemini31pro2026} supports a context length of 1 million tokens while Llama 4 Scout~\cite{meta_llama4scout} can support up to 10 million tokens.

Scaling context window size 
is challenging because
the computation and memory requirements
of attention---the core component of LLMs---increase
drastically with the increasing context window size.
For this reason, {\em distributed attention}
is 
intensively researched to ensure that long context
window can leverage increasing amount of
computation and memory resource
\rev{
of multiple 
AI accelerators (e.g., GPUs).
}

\begin{figure*}[htbp]
    \centering
    \includegraphics[width=\linewidth]{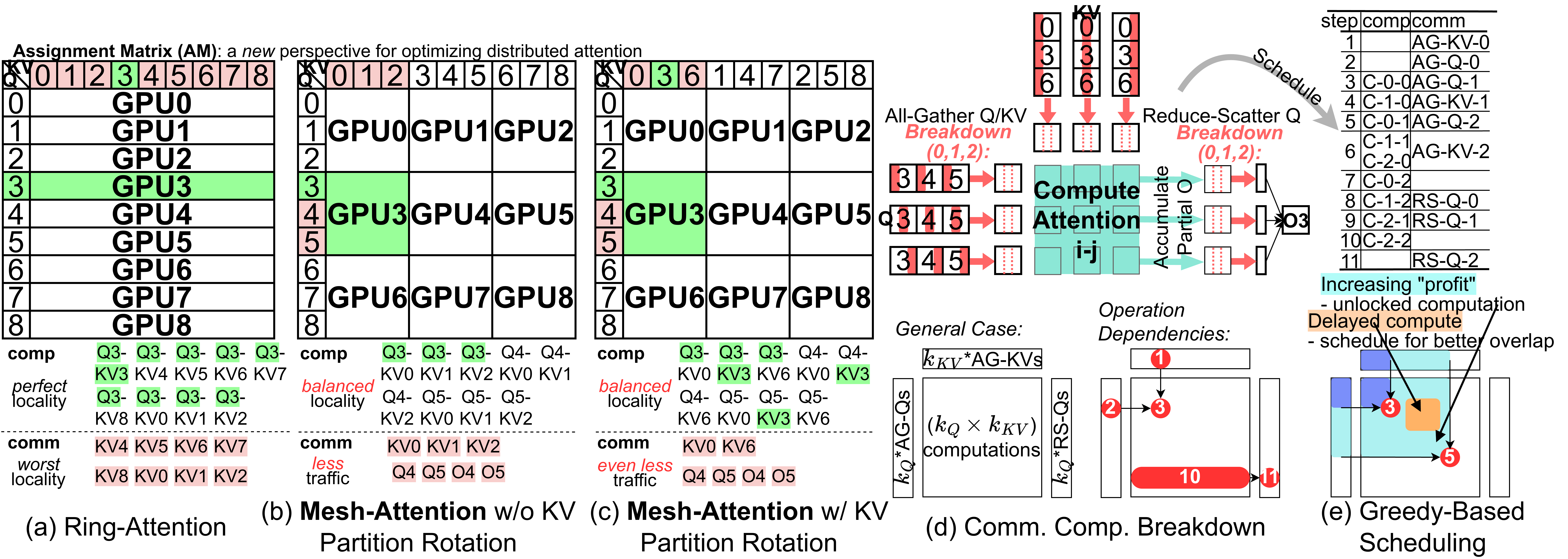}
    \vspace{-6mm}
  \caption{From Ring-Attention to {\em \proj}
  }
  \vspace{-5mm}
  \label{fig:overview}
\end{figure*}

\rev{
However, existing state-of-the-art distributed attention approaches either support only a limited degree of parallelism or incur high communication overhead. Ulysses~\cite{jacobs2023deepspeedulyssesoptimizationsenabling} uses efficient all-to-all collectives for communication, but its head-parallel design limits the parallelism degree to the number of attention heads. In contrast, Ring-Attention~\cite{liu2023ringattentionblockwisetransformers} can scale beyond this head-count limit, but incurs excessive communication cost.
}

\rev{
We introduce a Q-KV interaction matrix, called the \emph{assignment matrix} (AM), to analyze the communication overhead of Ring-Attention. Consider non-causal distributed attention on $n$ GPUs. The Q and KV tensors are each divided into $n$ partitions along the sequence dimension, with each GPU holding one Q partition and one KV partition. Computing global attention requires all $n^2$ interactions between the Q and KV partitions, while the final output partition for each Q partition must reside on its owning GPU. The AM is an $n \times n$ matrix whose $(i,j)$-th entry identifies the GPU assigned to compute the interaction between Q partition $i$ and KV partition $j$.
}

\rev{
Figure~\ref{fig:overview}(a) shows an assignment matrix for Ring-Attention on nine GPUs. Ring-Attention uses a Q-stationary, KV-moving method. Each GPU keeps its local Q partition and computes its interactions with all KV partitions. Thus, each GPU is assigned one row of the assignment matrix. To complete this row, each GPU must receive every remote KV partition. 
This row-wise assignment allows Ring-Attention to overlap communication with computation. It transfers the KV partitions through a logical ring while keeping each Q partition on its owning GPU. In this nine-GPU example, the execution consists of nine steps. At each step, every GPU computes the interaction between its local Q partition and the KV partition it currently holds, while the KV partitions move to the next GPUs in the ring at the same time. After nine steps, each GPU has processed all KV partitions and obtained the output partition corresponding to its local Q partition.
}

\rev{
The AM makes the communication cost of Ring-Attention explicit.
In Figure~\ref{fig:overview}(a), each GPU owns one row of the matrix,
representing the interactions between its local Q partition and all nine
KV partitions.
One KV partition is local, while the other eight reside on remote GPUs.
Under multi-head attention, each Q, K, and V partition has the same size,
which we use as one unit.
Because each KV partition contains both K and V,
each GPU receives $2 \times 8 = 16$ units of data while computing nine
Q-KV pairs\footnote{
A column-wise assignment initially communicates only Q and thus reduces
input communication.
However, it must subsequently return and merge the partial outputs for
each Q partition.
Recent work explores hybrid row/column schedules that select the assignment
based on operand sizes~\cite{yang2025contextparallelismscalablemilliontoken}.
}.
}

\rev{
We argue that Ring-Attention's communication inefficiency stems from an imbalance between its Q- and KV-dimension locality.
It maximizes Q locality by keeping each GPU's Q partition local, but minimizes KV locality by requiring every GPU to fetch all remote KV partitions.
Ring-Attention therefore occupies an extreme point in the locality trade-off that does not minimize total data movement.
By prioritizing Q locality at the expense of KV locality, it incurs a per-GPU communication volume that grows linearly with sequence length.
For example, with a hidden size of 8192, a sequence length of one million tokens, and BF16 data, each GPU communicates approximately 32 GiB of KV data per layer.
On 256 GPUs, critical-path communication accounts for 89\% and 93\% of causal attention forward and backward runtime, respectively, indicating severe communication bottleneck.
}

\rev{
To overcome the limitation, we propose \emph{\proj}, a distributed attention approach that {\em balances Q and KV locality} to reduce total communication.
Rather than assigning each GPU a one-dimensional row or column, \proj assigns it {\em a tunable two-dimensional tile}.
Varying the tile shape allows \proj to balance locality along both dimensions and minimize the global communication cost.
Figure~\ref{fig:overview}(b) illustrates one possible assignment using $3 \times 3$ tiles.
The tiles need not be square, but all GPUs receive tiles of the same shape and therefore perform the same amount of computation.
In this assignment, each diagonal GPU (i.e., GPUs 0, 4, and 8) communicates $2 + 2 \times 2 = 6$ units of Q and KV data, while each off-diagonal GPU communicates $2 + 3 \times 2 = 8$ units.
The partial outputs must then be aggregated across the three GPUs in each tile row.
Each GPU sends one output partition to each of its two peers, adding two units of P2P communication per GPU.
For example, GPU 0 sends its partial outputs to GPUs 1 and 2.
The aggregate communication volume is therefore $6 \times 3 + 8 \times 6 + 2 \times 9 = 84$ units, compared with $16 \times 9 = 144$ units for Ring-Attention.
}

\rev{
The naive 2-D tiling produces an irregular communication pattern because GPUs on the diagonal communicate less data than off-diagonal GPUs.
More importantly, its KV-dimension communication cannot be expressed using library-optimized collectives such as all-gather and must instead rely on less efficient P2P transfers.
To address this problem, we introduce \emph{KV Partition Rotation (KVR)}, a simple technique that permutes the KV partition IDs in the assignment matrix to regularize and balance communication.
Figure~\ref{fig:overview}(c) illustrates an example that changes the KV partition order from $(0,1,2,3,4,5,6,7,8)$ to $(0,3,6,1,4,7,2,5,8)$ such that the three indices for each GPU column are exactly the IDs of the GPUs
in the column.
After this permutation, every GPU communicates the same amount of data, and the major communication operations can be implemented using subgroup all-gather and reduce-scatter collectives.
In Figure~\ref{fig:overview}(c), every GPU, whether diagonal or off-diagonal, receives two Q partitions and two KV partitions and communicates two units for partial-output aggregation.
Because each KV partition contains both K and V, the aggregate communication volume becomes $(2 + 2 \times 2) \times 9 + 2 \times 9 = 72$ units, down from 84 units before KVR.
KVR is also applicable to non-square tiles,
which will be discussed in details in Section~\ref{sec:am_assign}.
}

\rev{
Reducing communication volume only addresses half of the problem---the remaining challenge is to hide communication behind computation.
Ring-Attention overlaps each KV transfer with computation on the current KV partition, but this fixed one-dimensional pipeline does not generalize to \proj's two-dimensional assignment.
To address this challenge, we introduce \emph{2-D Greedy Overlapping Optimization (GOO)}.
GOO partitions large all-gather, reduce-scatter, and attention computation kernels into smaller communication/computation tasks, allowing communication to overlap with independent computation that is ready to execute.
The resulting two-dimensional task graph, however, contains complex dependencies among Q communication, KV communication, attention computation, and output aggregation (Figure~\ref{fig:overview}(d)).
Our greedy scheduler tracks these dependencies and prioritizes the communication task that unlocks the most computation per unit of estimated communication time.
It then overlaps the selected communication task with computation that is already ready to execute (Figure~\ref{fig:overview}(e)).
}

\rev{
Modern GPU clusters typically have heterogeneous network topology. GPUs within a node communicate over fast scale-up links, whereas communication across nodes traverses slower scale-out links.
To exploit this hierarchy, we introduce \emph{Topology-Aware Mapping (TAM)}, which maps logical GPUs in the assignment matrix to physical GPUs
in the cluster
with network-topology awareness,
to reduce traffic over slower inter-node links.
}

\rev{
Combining these techniques, we implement \proj as a distributed attention library.
Our theoretical analysis in Section~\ref{sec:theoretical_scalability_analysis} shows that \proj reduces the asymptotic per-GPU communication complexity by a factor of $\sqrt{n}$ relative to one-dimensional schemes like Ring-Attention, where $n$ is the number of GPUs.
Across experiments with up to 256 GPUs and sequences of up to 1M tokens, \proj consistently outperforms all evaluated state-of-the-art methods.
Compared with Ring-Attention~\cite{liu2023ringattentionblockwisetransformers}, USP~\cite{fang2024uspunifiedsequenceparallelism}, and StarTrail~\cite{liu2025startrailconcentricringsequence}, \proj achieves maximum speedups of \attnmaxspeedupring, \attnmaxspeedupusp, and \attnmaxspeedupstartrail, and geometric-mean speedups of \attnavgspeedupring, \attnavgspeedupusp, and \attnavgspeedupstartrail, respectively.
\proj also demonstrates near-linear strong scaling and stable weak-scaling performance as the GPU count and sequence length increase.
}

%% file: background.tex
\section{Background}
\subsection{LLMs and Attention}

\rev{
Mainstream modern LLMs are based on the Transformer architecture~\cite{zhao2025surveylargelanguagemodels,vaswani2023attentionneed}.
Each Transformer layer contains a self-attention block followed by a position-wise feed-forward network.
The attention operation is defined as:
\begin{equation*}
    O = \operatorname{softmax}\left(\frac{QK^T}{\sqrt{d}}\right)V,
\end{equation*}
where $d$ is the attention head dimension.
}

\rev{
The matrix $QK^T$ contains the pairwise interactions between all query and KV tokens, and each row of its softmax weights determines how one query aggregates the corresponding value vectors.
This all-pairs structure naturally forms a two-dimensional Q-KV interaction space.
The attention computation scales quadratically and therefore dominates for sufficiently long sequences.
}

\rev{
FlashAttention~\cite{dao2022flashattentionfastmemoryefficientexact,dao2023flashattention2fasterattentionbetter} 
is a high-performance kernel that
tiles and fuses the attention operations to avoid materializing the quadratic attention matrix, substantially reducing memory traffic and intermediate storage.
However, it does not change the quadratic arithmetic complexity, and the QKV activation footprint still grows linearly with sequence length.
As LLM contexts expand to thousands or millions of tokens for long-context training and inference~\cite{TheC3,bai2024longbenchbilingualmultitaskbenchmark}, both costs make single-GPU attention increasingly impractical.
}

\rev{
To alleviate these bottlenecks, a variety of distributed attention schemes have been proposed
to allow multiple GPUs to 
process the attention of 
one single long sequence.
Distributed attention exploits 
{\em sequence parallelism},
a new parallelism that is additional to 
data parallelism~\cite{goyal2018accuratelargeminibatchsgd}, tensor parallelism~\cite{shoeybi2020megatronlmtrainingmultibillionparameter}, and pipeline parallelism~\cite{huang2019gpipeefficienttraininggiant,harlap2018pipedreamfastefficientpipeline}
for LLM training.
}

Sequence parallelism sharded the Q/K/V tensors sequence-wise to partitions and distributed to all GPUs. Consequently, cross-GPU communication is required to perform the core distributed attention operations in which all tokens interact along the Q-KV dimensions. Ultimately, each processor retains only the O partition corresponding to its assigned input region. 
Next, we discuss several recent distributed 
attention schemes.

\subsection{Ring-Attention}

\rev{
Ring-Attention~\cite{liu2023ringattentionblockwisetransformers} is a
\rev{commonly-used distributed attention algorithm}.
Each GPU keeps its local Q partition fixed while KV partitions circulate
through a logical Ring.
At each step, the GPU computes attention between its local Q partition
and the current KV partition and merges the  partial output into
a local accumulator.
The transfer of the next KV partition can proceed concurrently with the
attention computation on the current partition
to achieve communication-computation overlapping.
}

\rev{
Online softmax~\cite{milakov2018onlinenormalizercalculationsoftmax}
enables partial outputs computed over different KV partitions to
be merged exactly.
Rather than directly summing these partial outputs, online softmax
rescales them using their log-sum-exp (LSE) statistics before updating
the output accumulator.
}

Figure~\ref{fig:overview}(a) illustrates Ring-Attention on nine GPUs.
The Q, K, and V tensors are evenly split along the sequence
dimension into nine partitions, denoted by $\{Q_i\}_{i=0}^{8}$ and
$\{KV_i\}_{i=0}^{8}$, where $KV_i=(K_i,V_i)$.
Initially, GPU $i$ holds $Q_i$ and $KV_i$.
Over nine compute steps and eight K/V rotations, every GPU processes
each K/V partition exactly once while keeping its local Q partition stationary.
Each attention computation is expressed as
\begin{equation*}
    O_{i,j}, lse_{i,j} = \operatorname{Attention}(Q_i, K_j, V_j).
\end{equation*}
Here, $O_{i,j}$ is the partial output computed from $Q_i$ and $KV_j$,
and $lse_{i,j}$ contains the corresponding LSE statistics.
GPU $i$ incrementally merges all nine pairs
$\{(O_{i,j},lse_{i,j})\}_{j=0}^{8}$ into the final output $O_i$.
The final output $O_i$ has the same shape as $Q_i$ and is passed to the
subsequent model layer.
Because LSE stores only one scalar per query position and attention head, it
is much smaller than the corresponding output tensor.
Unless otherwise stated, we omit its contribution when discussing output
communication and partial-result merging throughout the remainder of this paper.

\subsection{Other Related Works}

\rev{
Yang et al.~\cite{yang2025contextparallelismscalablemilliontoken} extend
Ring-Attention with pass-KV and pass-Q variants for long-context inference.
Pass-Q circulates Q instead of KV and uses an additional All-to-All to collect
partial outputs before the final merge.
A runtime heuristic selects between the two variants according to the model
configuration and KV-cache hit rate.
}

\rev{
DeepSpeed Ulysses~\cite{jacobs2023deepspeedulyssesoptimizationsenabling}
uses All-to-All communication to transform sequence-sharded QKV tensors into
a head-sharded layout, computes complete attention heads locally, and then
restores the sequence-sharded output through a second All-to-All.
Its parallelism degree is limited by the number of attention heads.
USP~\cite{fang2024uspunifiedsequenceparallelism} combines Ulysses and
Ring-Attention in a two-dimensional process grid to scale beyond this limit.
}

\rev{
StarTrail~\cite{liu2025startrailconcentricringsequence} introduces an
additional team dimension on top of Ring-Attention.
It partitions $P$ GPUs into teams of size $C$ and first gathers QKV within
each team.
GPUs with the same local team rank then form  P2P sub-rings that
circulate larger team-level KV blocks.
This reduces the sub-ring degree to $P/C^2$ and lowers P2P communication
relative to a global Ring.
Finally, partial outputs are combined through a ReduceScatter within each
team.
This design provides a tunable trade-off between collective and P2P
communication, but only the sub-ring transfers overlap with blockwise
attention computation.
}

%% file: method.tex
\section{\proj}

\subsection{\rev{Definitions and Analysis}}

We formalize distributed attention using the assignment matrix introduced
in the Introduction.

\textbf{Assignment matrix (AM).}
Consider $n$ GPUs, with Q and KV evenly partitioned into
$\{Q_i\}_{i=0}^{n-1}$ and $\{KV_j\}_{j=0}^{n-1}$.
The assignment matrix $\mathbf{A}\in\{0,\ldots,n-1\}^{n\times n}$ assigns
each Q-KV interaction to a logical GPU: $A_{ij}=g$ means that GPU $g$
computes the interaction between $Q_i$ and $KV_j$.

\textbf{Q- and KV-dimension locality.}
The cells assigned to a GPU form its computation region in the AM.
We define the \emph{Q footprint} of this region as the set of distinct Q
partitions covered by it, and define the \emph{KV footprint} 
as the set of covered distinct KV partitions.
For GPU $g$, let $\mathcal{S}_g=\{(i,j)\mid A_{ij}=g\}$ denote its computation
region.
Its Q and KV footprints are
\[
    I_g^Q=\{i\mid \exists j:(i,j)\in\mathcal{S}_g\},
    \qquad
    I_g^{KV}=\{j\mid \exists i:(i,j)\in\mathcal{S}_g\}.
\]
These footprints are exactly the Q and KV partitions that GPU $g$ must access
to execute its assigned computation.
Because GPU $g$ initially stores $Q_g$ and $KV_g$, these local partitions do
not require communication.
We therefore define the \emph{remote footprints} as
\[
    R_g^Q=I_g^Q\setminus\{g\},
    \qquad
    R_g^{KV}=I_g^{KV}\setminus\{g\}.
\]
We measure locality by the size of the corresponding remote footprint:
a smaller $\lvert R_g^Q\rvert$ or $\lvert R_g^{KV}\rvert$ indicates higher
Q- or KV-dimension locality.
Under multi-head attention, a KV partition is twice the size of a Q partition
because it contains both K and V.
However, accessing a remote Q partition also requires returning a partial
output of the same size to its owner.
Ignoring the much smaller LSE statistics, a remote Q partition and a remote
KV partition therefore incur the same communication volume.
An efficient assignment should consequently keep
$\lvert R_g^Q\rvert+\lvert R_g^{KV}\rvert$
small for every GPU.

\textbf{Ring-Attention.}
Ring-Attention assigns row $i$ of the AM to GPU $i$.
Each GPU therefore accesses no remote Q partitions but all $n-1$ remote KV
partitions.
Thus, the total remote footprint of every GPU $i$ is
$\lvert R_i^Q\rvert+\lvert R_i^{KV}\rvert=n-1$.
Since every KV partition contains both K and V, this assignment communicates
$2(n-1)$ units per GPU.
Ring-Attention thus achieves perfect Q locality at the expense of the worst
possible KV locality.

\subsection{\rev{Basic Tiling Based Workload Distribution}}
\label{sec:am_assign}

We assign each GPU a 2-D tile instead of a 1-D row or column.
Because an $n\times n$ AM contains $n^2$ cells, balancing computation across
$n$ GPUs assigns exactly $n$ cells to each GPU.
Any factorization $n=a\times b$ therefore defines a valid tile with height
$a$ along the Q dimension and width $b$ along the KV dimension.
The AM is divided into $b$ tile rows and $a$ tile columns.

Without loss of generality, we number the logical GPUs according to the
row-major order of their assigned tiles.
GPU $g$ is assigned tile row $r=\lfloor g/a\rfloor$ and tile column
$c=g\bmod a$.
Its tile covers Q partitions from $ra$ to $(r+1)a-1$ and KV partitions from
$cb$ to $(c+1)b-1$.
This assignment always includes the local partition $Q_g$, giving
$\lvert R_g^Q\rvert=a-1$.
It may or may not include $KV_g$, giving
$\lvert R_g^{KV}\rvert=b-1$ or $b$.

The locality bound is minimized when $a$ and $b$ are as close as possible.
When $n$ is a perfect square, choosing $a=b=\sqrt{n}$ gives
\[
    \max_g
    \left(
    \lvert R_g^Q\rvert+\lvert R_g^{KV}\rvert
    \right)
    \leq 2\sqrt{n}-1,
\]
compared with $n-1$ for Ring-Attention.
Hence,
with balanced tile dimensions, \proj reduces the asymptotic per-GPU
communication complexity from $O(n)$ for Ring-Attention to $O(\sqrt{n})$.

Figure~\ref{fig:overview}(b) shows an example with $n=9$ and $a=b=3$.
GPU 3 covers $I_3^Q=\{3,4,5\}$ and
$I_3^{KV}=\{0,1,2\}$.
After excluding its local partitions,
$R_3^Q=\{4,5\}$ and
$R_3^{KV}=\{0,1,2\}$.
Its total remote footprint is $5$, compared with $8$ for Ring-Attention.

Algorithm~\ref{algorithm:basic_2d_attn} summarizes the resulting distributed
attention algorithm. We omit the backward pass for simplicity.
Each GPU fetches its required remote Q and KV partitions, computes one partial
output for each Q partition in its tile, and returns each partial output to the
owner of the corresponding Q partition.
Each owner then merges all partial outputs associated with its local Q
partition using online softmax.

For example, GPU 3 in Figure~\ref{fig:overview}(b) fetches $Q_4$, $Q_5$,
and $KV_0$, $KV_1$, and $KV_2$.
It computes partial outputs for $\{Q_3,Q_4,Q_5\}$ over
$\{KV_0,KV_1,KV_2\}$, retains the result for $Q_3$, and sends the other two
results to GPUs 4 and 5.
Finally, it merges its local result with the partial outputs for $Q_3$
received from GPUs 4 and 5 to produce $O_3$,
the final attention output for
its local partition.

\begin{algorithm}[t]
\footnotesize
\SetNoFillComment
\DontPrintSemicolon
\caption{Basic 2-D Mesh Distributed Attention for GPU $g$ (Forward)}
\label{algorithm:basic_2d_attn}
\SetKwInOut{Input}{Input}
\SetKwInOut{Output}{Output}

\Input{Local $Q_g$, $KV_g$, and assignment matrix $\mathbf{A}$}
\Output{Local output $O_g$}

Determine $I_g^Q$, $I_g^{KV}$, $R_g^Q$, and $R_g^{KV}$ from $\mathbf{A}$\;

Fetch $\{Q_i\mid i\in R_g^Q\}$ and
$\{KV_j\mid j\in R_g^{KV}\}$ from their owners\;

\ForEach{$Q_u$, $KV_v$ where $u\in I_g^Q$ and $v\in I_g^{KV}$}{
    Compute attention between $Q_u$ and $KV_v$\;
    Update partial output $\widetilde{O}_{g,u}$ using the online-softmax reducer\;
}

Send each $\widetilde{O}_{g,u}$ to the owner of $Q_u$\;

Receive all partial outputs associated with $Q_g$\;

Merge them using the online-softmax reducer to produce $O_g$\;
\end{algorithm}

\subsection{\rev{KV Partition Rotation in the Assignment Matrix}}

A major limitation of Algorithm~\ref{algorithm:basic_2d_attn} is its
irregular and imbalanced KV communication pattern.
The KV footprint $I_g^{KV}$ of GPU $g$ may or may not contain its local
partition $KV_g$.
Consequently, its remote footprint contains either $b-1$ or $b$ KV
partitions, causing different GPUs to communicate different amounts of data.
In the nine-GPU example, the diagonal GPUs fetch two remote KV partitions,
whereas the remaining GPUs fetch three.
More importantly, the owners of the required KV partitions do not align with
the GPUs that consume them.
\begin{wrapfigure}{r}{0.33\columnwidth}
    \centering
    \includegraphics[width=\linewidth]{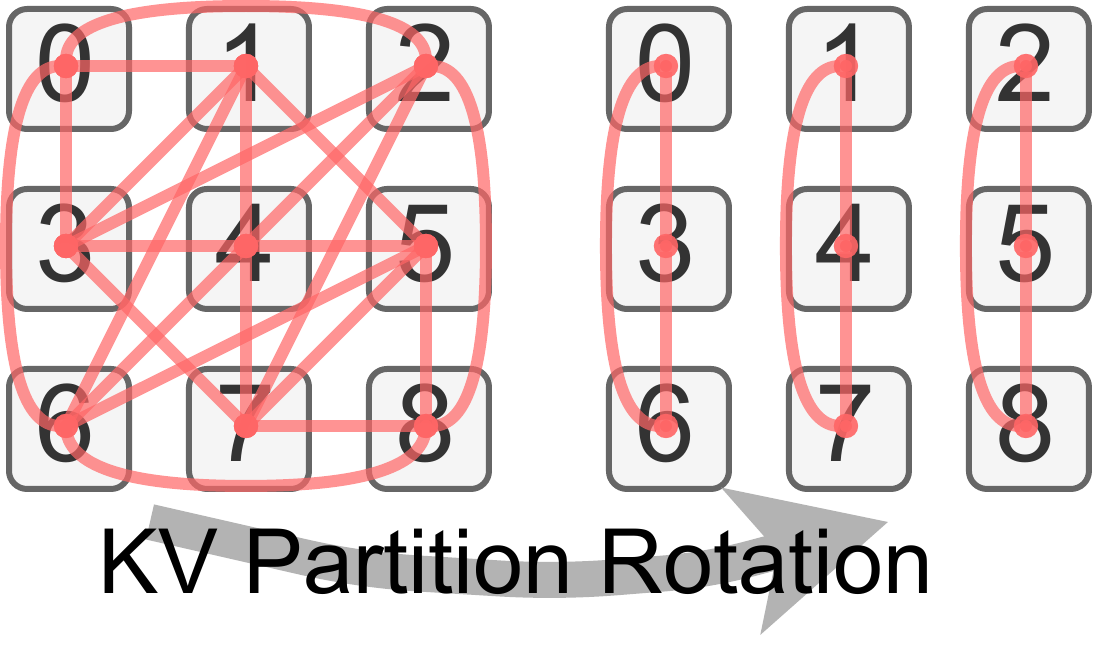}
    \vspace{-6mm}
  \caption{KV transfer pattern before / after KV partition rotation}
  \label{fig:kv_rotation_partition}
\end{wrapfigure}
For example, one tile column may require $KV_0$, $KV_1$, and $KV_2$, while
the corresponding computation is assigned to GPUs 0, 3, and 6.
This redistribution cannot be expressed as a regular subgroup all-gather.
Instead, it requires an irregular all-to-all that is typically decomposed
into multiple P2P transfers.
The left panel of Figure~\ref{fig:kv_rotation_partition} illustrates the
resulting nonuniform communication pattern.

One could implement a specialized communication primitive for this pattern,
but doing so would require substantial engineering effort and could not
directly leverage the highly optimized collectives provided by modern GPU
communication libraries.
Instead, we introduce \emph{KV Partition Rotation (KVR)}, which permutes the
KV partition IDs without changing the tile assignment or computation load.

Before rotation, tile column $c$ covers the contiguous KV partitions from
$cb$ to $(c+1)b-1$.
KVR replaces them with the strided KV footprint
\[
    I_g^{KV}=\{c+ta\mid 0\leq t<b\},
    \qquad c=g\bmod a.
\]
The GPUs in tile column $c$ have exactly the same IDs,
$\{c,c+a,\ldots,c+(b-1)a\}$.
Therefore, they collectively own all KV partitions required by that tile
column.
For example, GPU 3 in Figure~\ref{fig:overview}(c) belongs to tile column
$c=0$.
After KVR, its KV footprint becomes $\{0,3,6\}$ instead of $\{0,1,2\}$.
Because GPU 3 already owns $KV_3$, it fetches only $KV_0$ and $KV_6$.
More generally, KVR ensures that every GPU's KV footprint includes its local
KV partition, so all GPUs fetch exactly $b-1$ remote KV partitions.
Because the GPUs in each tile column collectively own its complete KV
footprint, the redistribution becomes an all-gather within that group.

With KVR, communication becomes regular and group-local. Every data exchange
is confined to a tile row or tile column.
Q partitions and partial outputs are communicated among GPUs in the same
tile row, whereas KV partitions are communicated among GPUs in the same tile
column.
The original $n$-GPU communication pattern can therefore be decomposed into
a 2-D mesh containing two types of communication groups.

\textbf{Q groups.}
Tile row $r$ forms a Q group
$G_r^Q=\{ra+x\mid 0\leq x<a\}$, where $0\leq r<b$.
Each Q group contains the $a$ consecutive GPUs in the same tile row.

\textbf{KV groups.}
Tile column $c$ forms a KV group
$G_c^{KV}=\{c+ax\mid 0\leq x<b\}$, where $0\leq c<a$.
Each KV group contains the $b$ GPUs in the same tile column.

\begin{algorithm}[t]
\footnotesize
\SetNoFillComment
\DontPrintSemicolon
\caption{2-D Mesh Distributed Attention with KV Partition Rotation for GPU $g$ (Forward)}
\label{algorithm:2d_attn_kvr}
\SetKwInOut{Input}{Input}
\SetKwInOut{Output}{Output}

\Input{Local partitions $Q_g$ and $KV_g$}
\Output{Final local output partition $O_g$}

All-gather Q partitions within GPU $g$'s Q group\;
All-gather KV partitions within GPU $g$'s KV group\;

\ForEach{Q-KV pair $(Q_u,KV_v)$ in the gathered partitions}{
    Compute the partial attention result for $(Q_u,KV_v)$\;
    Merge the result into local $O_u$ using the online-softmax reducer\;
}

Reduce-scatter the partial $O$ partitions within GPU $g$'s Q group using the
online-softmax reducer\;
\end{algorithm}

Each GPU $g$ belongs to exactly one Q group,
$G_{\lfloor g/a\rfloor}^Q$, and one KV group,
$G_{g\bmod a}^{KV}$.
The communication of 2-D tiled attention can be decomposed into
a few coarse-grained collectives within the Q and KV groups
as shown in Algorithm~\ref{algorithm:2d_attn_kvr}.
For each GPU, the required Q and KV partitions are collected through two
all-gather operations within its Q and KV groups, respectively (line 1-2).
The GPU then computes the attention result for every Q-KV pair in its tile (line 3-5).
Finally, partial outputs associated with the same Q partition are combined
through reduce-scatter within the Q group (line 6).
We use online softmax~\cite{milakov2018onlinenormalizercalculationsoftmax} as a reducer over partial attention results.

\begin{figure}[htbp]
\vspace{-3mm}
    \centering
    \includegraphics[width=\linewidth]{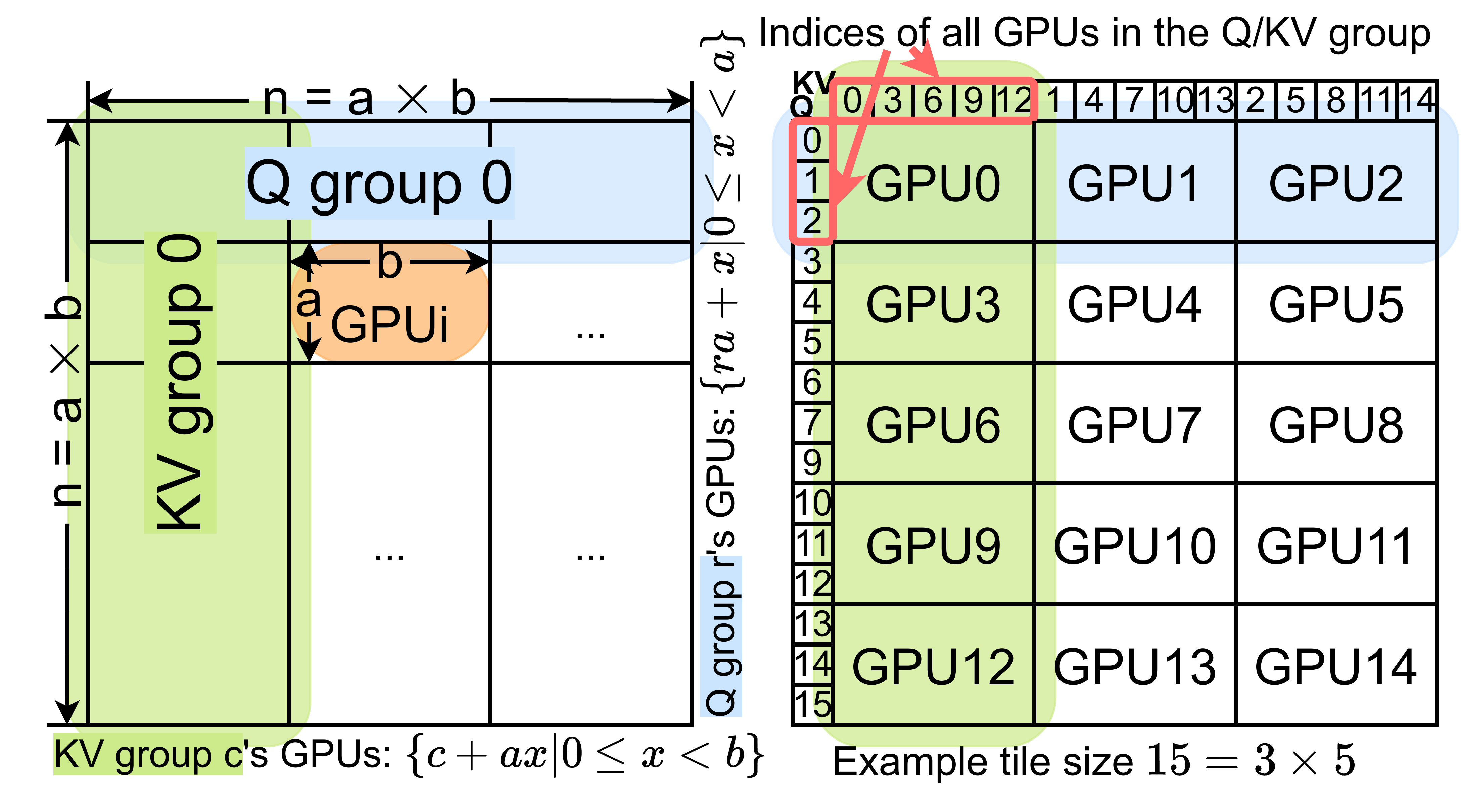}
    \vspace{-6mm}
  \caption{General \proj workload}
  \vspace{-1mm}
  \label{fig:general_mesh}
\end{figure}

The right panel of Figure~\ref{fig:general_mesh} shows an example with
$a=3$, $b=5$, and 15 GPUs.
GPU 0 belongs to Q group $G_0^Q=\{0,1,2\}$ and KV group
$G_0^{KV}=\{0,3,6,9,12\}$.
It uses two all-gather operations to collect
$\{Q_0,Q_1,Q_2\}$ and
$\{KV_0,KV_3,KV_6,KV_9,KV_{12}\}$,
and computes a partial output for each gathered Q partition over these KV
partitions.
GPUs 1 and 2 compute partial outputs for the same Q partitions over their
respective KV groups,
$\{KV_1,KV_4,\ldots,KV_{13}\}$ and
$\{KV_2,KV_5,\ldots,KV_{14}\}$.
Together, the three GPUs cover all 15 KV partitions.
They then perform a reduce-scatter within Q group $G_0^Q$ using online
softmax as the reducer.
This operation merges the partial outputs and returns $O_g$ to GPU $g$,
where $O_g$ is the exact attention output of $Q_g$ over all KV partitions.

\subsection{\rev{2-D Greedy Overlapping Optimization}}\label{sec:overlapping}

\begin{figure*}[t]
    \centering
    \includegraphics[width=\linewidth]{figures/greedy-overlapping-microchunks-breakdown.pdf}
    \caption{Overlapped \proj with collective communication breakdown}
  \vspace{-5mm}
  \label{fig:all_gather_breakdown}
  \vspace{-1mm}
\end{figure*}

The next challenge is to overlap communication with computation and hide as
much communication overhead as possible.
A natural strawman approach is to borrow the one-dimensional pipeline of
Ring-Attention by decomposing either the Q or KV all-gather into multiple
P2P transfers.
This produces two Ring-based designs.
1)
\textit{AGQ-RingKV:}
This design retains the Q-dimension all-gather but replaces the KV-dimension
all-gather with a Ring.
Within each KV group, GPUs overlap attention computation on the current KV
partition with its transfer to the next GPU.
The resulting partial outputs are then combined through reduce-scatter within
each Q group.
2)
\textit{AGKV-RingQ/O.}
This design retains the KV-dimension all-gather and circulates each Q
partition together with its partial output within the Q group.
Because every hop depends on the output updated by the previous hop, this
design has a sequential dependency that limits overlap.

Both designs optimize only one communication dimension at a time and cannot
coordinate overlap across the Q and KV dimensions.
To hide communication along both dimensions, we introduce
\emph{2-D Greedy Overlapping Optimization (GOO)}.
GOO decomposes each Q and KV collective into a small number of smaller
collective tasks, decomposes the corresponding attention computation at a
matching granularity, and greedily schedules these tasks to increase
communication-computation overlap.

Specifically, GOO decomposes the all-gather and reduce-scatter operations in
Algorithm~\ref{algorithm:2d_attn_kvr} along the sequence dimension.
Each local Q partition is divided into $k_Q$ chunks, while each local KV
partition is divided into $k_{KV}$ chunks.
The original Q all-gather and output reduce-scatter are consequently
decomposed into $k_Q$ smaller collectives, and the KV all-gather is decomposed
into $k_{KV}$ smaller collectives.
Each chunk-level collective becomes an independently schedulable
communication task.

Figure~\ref{fig:all_gather_breakdown} illustrates this decomposition.
The example contains 16 GPUs, each holding a local partition of 12 tokens.
GPU 4 holds tokens 48--59, while the other GPUs in its Q group,
GPUs 5, 6, and 7, hold tokens 60--71, 72--83, and 84--95, respectively.
A conventional Q-group all-gather materializes all tokens 48--95 before
computation can begin.

With $k_Q=3$, each local Q partition is divided into three four-token chunks.
For GPU 4, these chunks contain tokens 48--51, 52--55, and 56--59.
The original all-gather is then replaced by three smaller collectives:
AG-Q-0, AG-Q-1, and AG-Q-2.
AG-Q-0 gathers the first chunk from every GPU in the Q group, producing a
concatenated Q tensor containing tokens 48--51, 60--63, 72--75, and 84--87.
AG-Q-1 and AG-Q-2 similarly gather the second and third chunks, respectively.
The KV all-gather and output reduce-scatter are decomposed analogously into
$k_{KV}$ AG-KV tasks and $k_Q$ RS-Q tasks.

The communication decomposition divides the attention tile into
$k_Qk_{KV}$ compute tasks.
We denote the attention computation between the outputs of AG-Q-$i$ and
AG-KV-$j$ as C-$i$-$j$, where $0\leq i<k_Q$ and
$0\leq j<k_{KV}$.
GOO must therefore schedule $2k_Q+k_{KV}$ communication tasks together with
$k_Qk_{KV}$ compute tasks.
Based on empirical measurements, we restrict $k_Qk_{KV}\leq16$ to avoid
overly fine-grained tasks and preserve the efficiency of both attention
kernels and communication collectives.

The remaining challenge is to schedule these tasks while respecting their
two-dimensional dependencies.
A compute task C-$i$-$j$ becomes ready only after both AG-Q-$i$ and
AG-KV-$j$ complete.
Similarly, RS-Q-$i$ becomes ready only after all $k_{KV}$ compute tasks
C-$i$-$j$ associated with Q chunk $i$ complete.
Because these dependencies span both dimensions, the amount of ready
computation is determined jointly by the entire sequence of scheduling
decisions rather than by any single communication task.
An imbalanced schedule may leave too little computation ready for overlap
during part of the execution, while releasing a large compute backlog only
near the end, after most communication has completed.
The scheduler must therefore coordinate progress along both dimensions to
maintain sufficient ready computation and avoid both exposed communication
and a long un-overlapped computation tail.

To address this scheduling problem, we propose a step-based greedy scheduler.
The scheduler creates $2k_Q+k_{KV}$ execution steps, one for each
chunk-level collective.
Each step launches exactly one communication task together with zero or more
compute tasks that execute concurrently.
The step semantics are strict. All tasks assigned to a step must be ready when the step begins, and both communication and computation
must finish before the next step starts.
The scheduler follows two heuristics.

\textbf{Unlocking more computation at lower cost.}
The scheduler should maintain sufficient ready computation throughout
execution so that later communication can also be hidden.
For each candidate communication task $c$, let $T_{\mathrm{comm}}(c)$ denote its
estimated communication time and let $T_{\mathrm{unlock}}(c)$ denote the
estimated execution time of the compute tasks that would become newly ready
under the current schedule state.
We define its profit as
\[
    \operatorname{profit}(c)
    =
    \frac{T_{\mathrm{unlock}}(c)}
         {T_{\mathrm{comm}}(c)}.
\]
The scheduler greedily selects the candidate with the highest profit,
favoring communication that creates more future overlap opportunities at
lower cost.
Chunks belonging to the same collective type have identical message sizes,
so we profile AG-Q, AG-KV, and RS-Q once each and reuse their measured costs.

\textbf{Executing \emph{just enough} computation.}
After selecting a communication task, the scheduler fills the step using
compute tasks that were already ready at the step boundary.
It adds compute tasks until their total estimated execution time reaches
\[
    T_{\mathrm{target}}(c)
    =
    \gamma T_{\mathrm{comm}}(c),
\]
where $\gamma$ is a small safety factor.
Empirically,
we choose $\gamma=1.05$.
Executing less computation would leave part of the communication exposed,
whereas executing substantially more would delay the next collective without
providing additional overlap.
The remaining ready tasks are deferred to later steps, preserving computation
that can hide subsequent communication.
Within this budget, the scheduler favors computation tasks that complete an earlier Q
chunk, allowing its RS-Q task to become ready sooner.

We illustrate the greedy scheduling process using the example in
Figure~\ref{fig:all_gather_breakdown}. Assume that AG-KV, AG-Q, RS-Q, and
C tasks take 1, 0.5, 0.5, and 0.5 time units, respectively. For clarity,
we set $\gamma=1$ in this example and consider only the lowest-index
unscheduled chunk of each collective type. Ties are broken arbitrarily.
Initially, neither AG-KV-0 nor AG-Q-0 can unlock a compute task alone, so both
have zero profit. We choose AG-KV-0 for the first step. After it completes,
AG-Q-0 unlocks C-0-0 and has a profit of $0.5/0.5=1$, whereas AG-KV-1 still
has zero profit. We therefore schedule AG-Q-0 in step 2.
In step 3, AG-KV-1 and AG-Q-1 each unlock one compute task. Because AG-Q-1
takes only half as long as AG-KV-1, its profit is twice as high. We therefore
schedule AG-Q-1 together with the already-ready C-0-0. Applying the same rule,
the scheduler selects AG-KV-1, AG-Q-2, and AG-KV-2 for steps 4--6,
respectively. At the beginning of step 6, C-1-1 and C-2-0 are ready. Their
combined execution time matches that of AG-KV-2, so the scheduler executes
both concurrently with AG-KV-2, fully hiding its communication.

It is worth noting that all GPUs must follow the same communication schedule. Otherwise, they may
invoke collectives in different orders and deadlock. 
\rev{
We therefore profile task costs on all GPUs, average the measurements across
them, and run the same deterministic greedy planner on every GPU using these
shared costs.
}
This produces identical
schedules and prevents deadlocks caused by inconsistent collective ordering.
\rev{
Another corner case arises when no communication task is schedulable but
ready computation remains. We then insert a compute-only step with one ready
computation task to advance dependencies, as illustrated by steps 7 and 10 in
Figure~\ref{fig:all_gather_breakdown}.
}

\subsection{\rev{Backward Pass of \proj}}

The backward pass mirrors the 2-D dataflow of the forward pass.
GPU $g$ initially holds its local $Q_g$, $KV_g$, output $O_g$, and output
gradient $\mathrm{d}O_g$, where $\mathrm{d}O$, $\mathrm{d}Q$, and $\mathrm{d}KV$ denote the gradients of $O$, $Q$,
and $KV$, respectively. Its goal is to produce the local gradients $\mathrm{d}Q_g$ and
$\mathrm{d}KV_g$. Algorithm~\ref{algorithm:2d_kvr_bwd} summarizes this process.

Each GPU first all-gathers the Q, O, and dO partitions within its Q group and
all-gathers the KV partitions within its KV group. For each gathered pair
$(Q_i,KV_j)$, it invokes the backward attention kernel using
$(Q_i,KV_j,O_i,\mathrm{d}O_i)$ to compute partial gradients for $Q_i$ and $KV_j$.
These partial gradients are accumulated locally in $\widetilde{\mathrm{d}Q}_i$ and
$\widetilde{\mathrm{d}KV}_j$. After processing all pairs, the GPUs in each Q group
reduce-scatter their partial dQ gradients to produce $\mathrm{d}Q_g$, while the GPUs
in each KV group reduce-scatter their partial dKV gradients to produce
$\mathrm{d}KV_g$. Both reduce-scatter operations use addition as the reducer.

The backward pass also consists of multiple all-gather and reduce-scatter
collectives. We apply GOO in the same way---it decomposes these collectives and
the corresponding backward computation into smaller tasks, then uses the same
greedy heuristics to overlap communication with computation. 
We omit the
details of the
resulting backward schedule for brevity.

\begin{algorithm}[t]
\footnotesize
\SetNoFillComment
\DontPrintSemicolon
\caption{2-D Mesh Distributed Attention with KV Partition Rotation for GPU $g$ (Backward)}
\label{algorithm:2d_kvr_bwd}
\SetKwInOut{Input}{Input}
\SetKwInOut{Output}{Output}

\Input{Local $Q_g$, $KV_g$, $O_g$, and $\mathrm{d}O_g$}
\Output{Local gradients $\mathrm{d}Q_g$ and $\mathrm{d}KV_g$}

All-gather $Q$, $O$, and $\mathrm{d}O$ partitions within GPU $g$'s Q group\;
All-gather KV partitions within GPU $g$'s KV group\;

\ForEach{gathered Q partition $Q_u$ and KV partition $KV_v$}{
    Compute partial gradients for $Q_u$ and $KV_v$ using
    $(Q_u,KV_v,O_u,\mathrm{d}O_u)$\;
    Accumulate them into local $\widetilde{\mathrm{d}Q}_u$ and
    $\widetilde{\mathrm{d}KV}_v$\;
}

Reduce-scatter the accumulated dQ partitions within GPU $g$'s Q group using summation\;
Reduce-scatter the accumulated dKV partitions within GPU $g$'s KV group using summation\;
\end{algorithm}

\subsection{\rev{Topology-Aware GPU Mapping}}

\begin{figure}[htbp]
    \centering
    \includegraphics[width=1.0\linewidth]{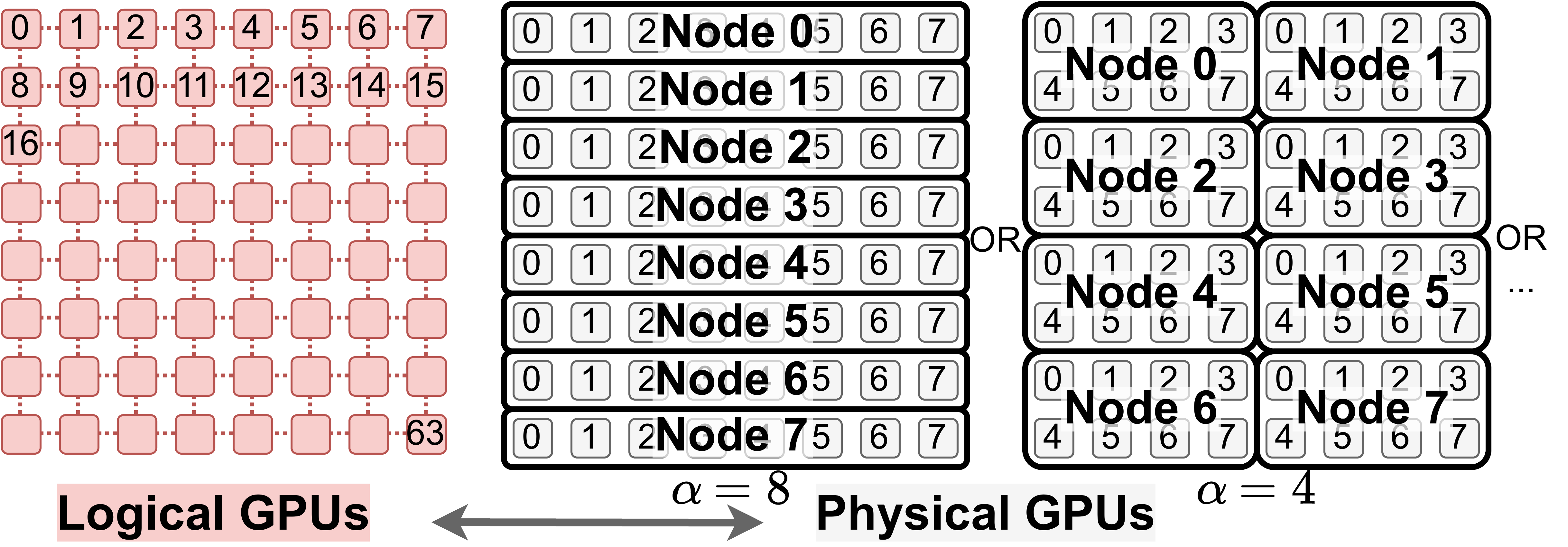}
    \vspace{-6mm}
  \caption{Logical--physical GPU mapping}
  \vspace{-2mm}
  \label{fig:logical_physical}
\end{figure}

Modern GPU clusters have hierarchical interconnects. GPUs within the same node
communicate through fast scale-up links, whereas communication across nodes
traverses slower scale-out links. Mapping the logical GPUs in the assignment
matrix onto physical GPUs is therefore non-trivial. The mapping should account
for the physical topology and minimize traffic over scale-out links.

A naive row-major mapping places consecutive logical GPUs on the same node.
By ignoring the 2-D group structure, it can favor locality along one dimension
over the other, resulting in imbalanced and excessive scale-out traffic.
Figure~\ref{fig:logical_physical} illustrates this issue using a 64-GPU
cluster with eight GPUs per node. Under the row-major mapping, logical GPUs
0--7 are mapped to node 0, while GPUs 8--15 are mapped to node 1. In this
particular configuration, each Q group is entirely contained within one node,
whereas every KV group spans all eight nodes. Consequently, all Q-dimension
communication is intra-node, while all KV-dimension communication traverses
scale-out links.
Assume that each Q, K, and V partition has unit size. Node 0 participates in
eight KV groups, each of which requires seven remote K and V partitions. Its
inter-node communication volume is therefore $2\times8\times7=112$ units,
where the factor of two accounts for K and V.

To reduce inter-node scale-out traffic along both dimensions, we instead map a 2-D tile
of the logical GPU mesh onto each node. As shown on the right of
Figure~\ref{fig:logical_physical}, node 0 is assigned logical GPUs 0--3 and
8--11 rather than GPUs 0--7. Node 0 now intersects two Q groups, each of which
exchanges four remote Q partitions and four corresponding output partitions.
The resulting Q-dimension traffic is $2\times2\times4=16$ units. It also
intersects four KV groups, each containing six remote KV partitions, resulting
in $2\times4\times6=48$ units of KV-dimension traffic. The total inter-node
communication is therefore 64 units, compared with 112 under the row-major
mapping.

We generalize this mapping using a tunable parameter $\alpha$, which specifies
the width of the logical 2-D tile assigned to each node. With eight GPUs per
node, the row-major mapping uses $\alpha=8$, whereas the improved mapping above
uses $\alpha=4$. In practice, we enumerate feasible values of $\alpha$ and
select the one with the best measured performance on the target cluster.

\subsection{\rev{Supporting Causal Masking}}

\rev{
Decoder-only LLMs such as GPT~\cite{brown2020languagemodelsfewshotlearners} use causal attention, where each token can attend only to itself and preceding tokens. Causal masking creates severe workload imbalance under contiguous sequence partitioning. For a pair $(Q_i, KV_j)$, the entire interaction is masked when $i<j$, triangular when $i=j$, and dense when $i>j$. Different Q-KV pairs therefore require substantially different amounts of computation. Since each 2-D tile contains multiple such pairs, tiles assigned to different GPUs can have highly imbalanced computational workloads.
To balance causal computation, we adopt a striped sequence layout similar to Striped Attention~\cite{brandon2023stripedattentionfasterring}. Rather than assigning contiguous sequence segments to GPUs, token positions are interleaved across the logical mesh so that each 2-D compute tile contains a similar fraction of unmasked Q-KV interactions. This preserves the 2-D communication structure and allows us to apply the same greedy scheduling framework.
}

\subsection{Theoretical Communication Complexity}\label{sec:theoretical_scalability_analysis}

\begin{table}[t]
  \centering
  \caption{Theoretical per-GPU forward communication volume under MHA.}
  \label{tab:theoretical_communication_volume}
  \renewcommand{\arraystretch}{1.25}
  \setlength{\tabcolsep}{2.5pt}
  \resizebox{\columnwidth}{!}{
  \begin{tabular}{@{}lccc@{}}
    \toprule
    \textbf{Method}
      & \textbf{Communication volume}
      & \textbf{Approximate minimum}
      & \textbf{SP constraint} \\
    \midrule
    Ring-Attention
      & $\left(2-\frac{2}{n}\right)Nd$
      & $2Nd$
      & No limit \\
    Ulysses
      & $\frac{4(n-1)}{n^2}Nd$
      & $\frac{4}{n}Nd$
      & $H$ \\
    StarTrail
      & $\left(\frac{4C-4}{n}+\frac{2}{C}\right)Nd$
      & $4\sqrt{\frac{2}{n}}Nd$
      & No limit \\
    \proj
      & $\left(\frac{2a+2b-4}{n}\right)Nd$
      & $\frac{4}{\sqrt{n}}Nd$
      & No limit \\
    \bottomrule
  \end{tabular}
  }
   \vspace{-3mm}
\end{table}

Table~\ref{tab:theoretical_communication_volume} compares the per-GPU forward
communication volume of \proj with Ring-Attention~\cite{liu2023ringattentionblockwisetransformers}, Ulysses~\cite{jacobs2023deepspeedulyssesoptimizationsenabling},
and StarTrail~\cite{liu2025startrailconcentricringsequence} under multi-head attention (MHA). We assume batch size one, sequence length
$N$, hidden dimension $d$, $H$ heads, and $n$ GPUs, and omit element size and
LSE metadata. Here, $C$ is StarTrail's team size ranging from $1$ to $\sqrt{n}$ and $ab=n$ defines \proj's
tile dimensions. We provide a brief derivation of these communication volumes below.

\begin{itemize}[wide, itemsep=0pt, parsep=0pt, partopsep=0pt]
\itemsep0em
    \item \textbf{Ring-Attention}. Each GPU sends \(n-1\) KV chunks, each of size \(2Nd/n\), to the successor in the ring, resulting in a communication volume of \((2-2/n)Nd\);
    \item \textbf{DeepSpeed Ulysses}. 
4 all-to-all operations of Q, K, V, and O chunks, each counted as \((n-1)/n^2Nd\). In total \(4(n-1)/n^2Nd\);
    \item \textbf{StarTrail}. The communication is divided into four stages: 
    an all-gather stage of \(3(C-1)dN/n\),
    a KV initialization stage of \(2CdN/n\),
    a P2P communication stage of \(2(n/C-C)Nd/n\),
    and a reduce-scatter stage of \((C-1)dN/n\).
    In total \(((4C-4)/n+2/C)Nd\).
  
    \item \textbf{\proj}. Each Q group of $a$ GPUs collectively performs 
AG-Q and RS-Q, while each KV group of $b$ GPUs
collectively performs AG-KV ($ab = n$). Every GPU holds
Q size $Nd/n$, KV size $2Nd/n$, and produces O size $Nd/n$. AG-Q therefore costs $(a-1)Nd/n$ and RS-Q costs the same, while the AG-KV costs
$2(b-1)Nd/n$. The volume in total:
\[
    2(a-1)\frac{Nd}{n} + 2(b-1)\frac{Nd}{n} = \frac{2(a+b-2)Nd}{n} .
\]
\end{itemize}

Ulysses achieves the lowest volume, $\Theta(Nd/n)$, but cannot use more than
$H$ GPUs.
Ring-Attention communicates $\Theta(Nd)$, whereas StarTrail and \proj require
$\Theta(Nd/\sqrt{n})$. With their optimal parameters, \proj uses a factor of
$\sqrt{2}$ less communication than StarTrail and a factor of
$\Theta(\sqrt{n})$ less than Ring-Attention. Backward preserves this
asymptotic ordering. 
The same asymptotic communication volume does not imply the same efficiency.
StarTrail's team all-gathers are not overlapped with attention computation,
while its Ring-based overlap relies on fine-grained P2P transfers. In
contrast, \proj overlaps coarse-grained collectives with attention computation
and uses topology-aware mapping to reduce inter-node traffic
(Section~\ref{sec:communication_analysis}).

%% file: experiment.tex
\section{Experiments}
\subsection{Experimental Settings}

\rev{
We evaluate \proj on a 256-GPU cluster unless stated otherwise. The cluster has 32 nodes, each equipped with eight GPUs and eight high-performance InfiniBand network adapters for inter-node GPU communication. Within each node, GPUs communicate over higher-bandwidth scale-up links.}
\rev{
We evaluate all experiments using 64 query heads with a head dimension of 128 (batch size: 1, data type: BF16). This attention geometry is representative of modern large language models: both Llama-3-70B~\cite{llama3modelcard} and Qwen3-32B~\cite{team2025qwen3} use 64 query heads with a head dimension of 128.
}

\rev{
We use four state-of-the-art distributed attention algorithms as baselines, including Ring-Attention~\cite{liu2023ringattentionblockwisetransformers}, Ulysses~\cite{jacobs2023deepspeedulyssesoptimizationsenabling}, USP~\cite{fang2024uspunifiedsequenceparallelism}, and StarTrail~\cite{liu2025startrailconcentricringsequence}.
We omit Ulysses for all results
with more than 64 GPUs due
to its parallelism limitation.
Because StarTrail has a complex design and no public implementation, we implement only its non-causal version. 
For StarTrail, we evaluate all valid team sizes and report the fastest result
for each setting. For \proj, we search over tile dimensions $(a,b)$ around the
communication-optimal balanced factorization, the Q- and KV-dimension chunk
counts $(k_Q,k_{KV})$ subject to $k_Qk_{KV}\leq16$, and the topology-aware
mapping parameter $\alpha$. We report the fastest configuration for each
setting. The search covers fewer than 200 configurations and takes from tens
of minutes to a few hours, depending on the workload and cluster scale. This
one-time tuning cost is negligible compared with large-scale LLM training,
which commonly runs for weeks or months.
For causal Ring-Attention and USP, we evaluate two variants, ZigZag~\cite{zhuzilin2024zigzag} and Stripe~\cite{brandon2023stripedattentionfasterring}, and report the better result.
By default, we set USP's Ulysses degree to 8, matching the eight GPUs per node. This is a common practice of USP that keeps each Ulysses all-to-all within a node while the Ring communication spans nodes.
}

\rev{
Throughout this section, NC and C denote non-causal and causal attention, while F and B denote forward and backward passes, respectively.
}

\newcommand{\rothead}[1]{%
  \makebox[0pt][c]{\rotatebox{90}{#1}}%
}

\subsection{Overall Performance}

\begin{table}[t]
\centering
\footnotesize
\caption{\rev{Normalized runtime and model FLOPs utilization (MFU) on 128 and 256 GPUs. Runtime is normalized to \proj under the
same setting.}}
\label{tab:combined_comparison}
\setlength{\tabcolsep}{1pt}
\scalebox{1.0}{%
\begin{tabular}{@{}l cccc cccc  @{}}
\toprule
& \multicolumn{4}{c}{\textbf{Norm.\ runtime}}
& \multicolumn{4}{c}{\textbf{MFU (\%)}} \\
\cmidrule(lr){2-5}\cmidrule(lr){6-9}
Setting & \projabbr & Ring & USP & StarTrail
        & \projabbr & Ring & USP & StarTrail \\
\midrule
\multicolumn{9}{@{}l}{\textbf{Forward}} \\
128G NC 256K & \textbf{1.0} & 16.5 & 2.3 & 4.0 & \textbf{49.1} & 3.0 & 21.3 & 12.3  \\
128G NC 512K & \textbf{1.0} &  9.4 & 1.8 & 2.5 & \textbf{55.1} & 5.9 & 31.0 & 22.5 \\
128G C 256K  & \textbf{1.0} & 29.3 & 3.8 & --  & \textbf{38.3} & 1.3 & 10.1 & --       \\
128G C 512K  & \textbf{1.0} & 17.7 & 2.3 & --  & \textbf{46.6} & 2.6 & 19.9 & --      \\
256G NC 512K & \textbf{1.0} & 20.0 & 2.6 & 3.4 & \textbf{50.0} & 2.5 & 19.3 & 14.8  \\
256G NC 1M   & \textbf{1.0} & 11.1 & 1.8 & 2.3 & \textbf{55.1} & 5.0 & 30.0 & 24.1  \\
256G C 512K  & \textbf{1.0} & 32.3 & 4.3 & --  & \textbf{42.5} & 1.3 & 10.0 & --      \\
256G C 1M    & \textbf{1.0} & 19.1 & 2.6 & --  & \textbf{50.4} & 2.6 & 19.7 & --     \\
\midrule
\multicolumn{9}{@{}l}{\textbf{Backward}} \\
128G NC 256K & \textbf{1.0} & 15.1 & 2.5 & 4.6 & \textbf{49.4} & 3.3 & 19.4 & 10.7 \\
128G NC 512K & \textbf{1.0} &  8.6 & 1.8 & 3.0 & \textbf{54.6} & 6.3 & 30.2 & 18.3\\
128G C 256K  & \textbf{1.0} & 27.7 & 3.9 & --  & \textbf{39.5} & 1.4 & 10.0 & --   \\
128G C 512K  & \textbf{1.0} & 16.0 & 2.5 & --  & \textbf{45.1} & 2.8 & 17.7 & --    \\
256G NC 512K & \textbf{1.0} & 16.4 & 2.6 & 2.7 & \textbf{44.7} & 2.7 & 17.1 & 16.3\\
256G NC 1M   & \textbf{1.0} &  9.4 & 1.8 & 2.0 & \textbf{50.3} & 5.3 & 27.4 & 25.7 \\
256G C 512K  & \textbf{1.0} & 30.2 & 4.4 & --  & \textbf{43.2} & 1.4 &  9.9 & --     \\
256G C 1M    & \textbf{1.0} & 16.7 & 2.7 & --  & \textbf{47.4} & 2.8 & 17.6 & --     \\
\bottomrule
\end{tabular}
}
\vspace{-2mm}
\end{table}

We compare the distributed-attention runtime and MFU~\footnote{All runtimes reported in this paper measure only the distributed attention operator and exclude QKV projections and other LLM components.} of \proj with Ring-Attention in Table~\ref{tab:combined_comparison}
with various sequence lengths and number of GPUs.
\rev{
\proj consistently outperforms Ring-Attention, USP and StarTrail in various settings and achieves a speedup of up to 32.3\(\times\), 4.4$\times$ and 4.6$\times$ (on average~\footnote{All reported average speedups are geomean in this paper.} 17.0$\times$, 2.6$\times$ and 2.9$\times$). 
\proj also consistently achieves a better MFU (model FLOPs utilization).
On average, \proj's MFU is 44.4, 28.2 and 32.9 percentage points (up to 50.1, 33.3 and 38.6) higher than Ring-Attention, USP and StarTrail, respectively. 
}
\rev{We note that for all algorithms,} MFU becomes higher as the sequence length increases. It is because the computation of attention increases 
quadratically w.r.t. sequence length while communication only increases linearly. 
Also, MFUs with causal mask are usually lower since causal mask reduces the computation by half, which makes communication a more severe bottleneck.

\subsection{Peak Memory Analysis}

We analyze the peak memory \rev{allocation} of \proj in Table~\ref{tab:peak_memory}.
\rev{
In general, 
\proj consumes more peak memory
than Ring-Attention and USP
and less memory than StarTrail.
The reason is that \proj 
}
needs to cache multiple KV/Q partitions throughout the entire attention computation
for data reuse whereas
Ring-Attention keeps at most 2 KV partitions and 1 Q partition in the memory.
However, it is worth noting that
the high peak memory consumption of \proj is transient:
most of the GPU memory will
be quickly released once the attention
computation of the current layer completes.
\proj will not increase the amount of stashed forward activations that are 
needed by the backward pass.
Also, as we will discuss in Section~\ref{sec:usp_experiments},
a USP variant of \proj 
can further reduce the peak memory by up to 73\% without sacrificing performance.
We leave exploring memory-efficient schedules (e.g., schedules that release KV/Q chunks in a more eager manner) as our future work.

\begin{table}[t]
\centering
\footnotesize
\caption{
\rev{Peak allocated memory.
Each result reports the maximum across all ranks.
Unit: GiB.}
}
\label{tab:peak_memory}

\begingroup
\setlength{\tabcolsep}{2pt}
\scalebox{1.0}{%
\begin{tabular}{@{}lcccccccc@{}}
\hline
\noalign{\vskip 1.5pt}
\multirow{2}{*}{Setting}
& \multicolumn{4}{c}{\textbf{Forward}}
& \multicolumn{4}{c}{\textbf{Backward}} \\[1pt]
\cline{2-5}\cline{6-9}
\noalign{\vskip 1.5pt}
& \projabbr & Ring & USP & StarTrail
& \projabbr & Ring & USP & StarTrail \\
\hline
128G NC 256K
& 1.8 & 0.6 & 0.7 & 3.9
& 3.6 & 0.9 & 1.0 & 7.6 \\
128G NC 512K
& 3.6 & 1.2 & 1.4 & 3.8
& 7.2 & 1.8 & 1.9 & 15.1 \\
128G C 256K
& 2.3 & 0.6 & 0.5 & --
& 3.7 & 0.9 & 0.9 & -- \\
128G C 512K
& 5.7 & 0.9 & 1.1 & --
& 10.4 & 1.4 & 1.8 & -- \\
256G NC 512K
& 2.4 & 0.6 & 0.7 & 3.6
& 5.0 & 0.9 & 1.0 & 14.9 \\
256G NC 1M
& 6.0 & 1.2 & 1.4 & 7.5
& 11.5 & 1.8 & 1.9 & 29.6 \\
256G C 512K
& 3.3 & 0.5 & 0.5 & --
& 4.8 & 0.9 & 1.0 & -- \\
256G C 1M
& 6.7 & 0.9 & 1.1 & --
& 11.9 & 1.4 & 1.8 & -- \\
\hline
\end{tabular}%
}
\endgroup
\vspace{-2mm}
\end{table}

\subsection{Scalability} 

\begin{figure}[!htbp]
 \vspace{-0mm}
    \centering
     \includegraphics[width=0.9\linewidth]{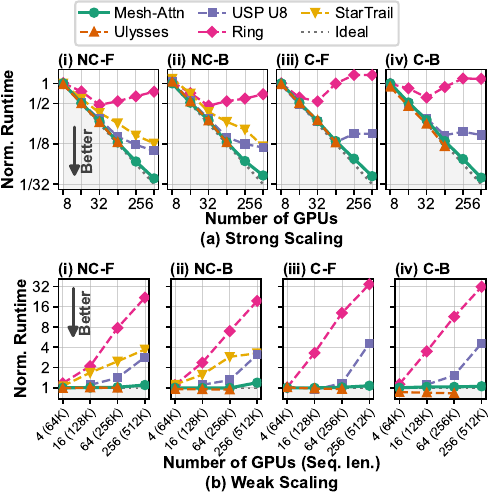}
  \caption{Strong scaling of sequence length 512K and weak scaling of sequence length from 64K to 512K. Normalized to the runtime of \proj at the smallest scale.}
  \vspace{-1mm}
  \label{fig:combined_scalability}
\end{figure}

\noindent\textbf{Strong Scalability.}
\rev{
Figure~\ref{fig:combined_scalability}(a) presents the strong-scaling results with a fixed sequence length of 512K tokens. Among the evaluated methods, \proj scales most effectively. Increasing the GPU count from 8 to 256 (a $32\times$ increase) reduces its runtime by a geometric mean of $25.10\times$ across the four evaluated methods. In comparison, Ring-Attention and USP achieve only $1.07\times$ and $7.22\times$ speedups, respectively, while StarTrail achieves $9.00\times$ across the two non-causal modes.
Ulysses also exhibits near-linear scaling up to 64 GPUs. However, its sequence-parallel degree must divide and cannot exceed the number of attention heads. With 64 heads, Ulysses therefore cannot scale to 128 or 256 GPUs.
}

\noindent\textbf{Weak Scalability.} \rev{
Under FLOP-balanced weak scaling (Figure~\ref{fig:combined_scalability}(b)), we increase the GPU count from 4 to 256 and the global sequence length from 64K to 512K, keeping the per-GPU attention FLOPs constant. The runtime of \proj increases by only $1.1\times$. In contrast, the runtimes of Ring-Attention and StarTrail increase by geometric means of $23.8\times$ and $3.2\times$, respectively, as communication increasingly dominates execution. USP U8 is not valid on 4 GPUs since its Ulysses degree is 8. 
From 16 GPUs to 256 GPUs, its runtime increases by $3.5\times$. Ulysses also exhibits near-ideal weak scaling through 64 GPUs, but its sequence-parallel degree is bounded by the number of attention heads.
}

\subsection{\proj-Based USP}
\label{sec:usp_experiments}


USP combines Ulysses all-to-all communication with a 
residual
 Ring-Attention lane.
 We construct \projabbr-USP by replacing this Ring lane with \proj. Tables~\ref{tab:usp-composability} compares \projabbr-USP, Ring-USP and standalone \proj for Ulysses degrees of 8, 16, and 32.

\projabbr-USP outperforms Ring-USP in 11 of the 12 settings, with speedups of up to $4.81\times$. The only exception is causal forward at U=32, where Ring-USP is $1.025\times$ faster. At this degree, the residual sequence-parallel degree is only $256/32=8$, which limits the benefit of replacing its Ring lane with \proj. \projabbr-USP also outperforms standalone \proj in all backward and causal settings. 
It is only $2.1\%-7.3\%$ slower in non-causal forward because it adds Ulysses all-to-all stages that are not overlapped with computation. Importantly, the Ulysses decomposition substantially reduces memory usage: at U=16, \projabbr-USP uses $2.67\times$ less peak memory on average (up to $2.99\times$), mitigating the memory overhead of standalone \proj.
This advantage comes from Ulysses' heads-to-sequence redistribution.
Each \proj lane in \projabbr-USP buffers multiple Q/KV sequence partitions for only a $\frac{1}{U}$ fraction of the attention heads, whereas standalone \proj maintains tile and communication state for all heads.

\begin{table}[t]
\centering
\footnotesize
\caption{
\rev{\projabbr-USP performance (normalized runtime) and peak allocated memory (GiB)
on 256 GPUs with a 512K sequence. Pure \projabbr and Ring-USP U=8 use their
corresponding performance-table configurations.}
}
\label{tab:usp-composability}
{\setlength{\tabcolsep}{2pt}
   \begin{tabular}{@{}l cccc @{\hspace{12pt}} cccc @{}}
\hline
& \multicolumn{4}{c}{Norm.\ runtime} & \multicolumn{4}{c}{Memory (GiB)} \\
\cline{2-5}\cline{6-9}
Method & NC-F & NC-B & C-F & C-B & NC-F & NC-B & C-F & C-B \\
\hline
Pure \projabbr
& 1.00 & 1.00 & 1.00 & 1.00 & 2.39 & 5.00 & 3.35 & 4.83 \\
Ring-USP, U=8
& 2.87 & 3.10 & 4.46 & 4.37 & 0.69 & 0.97 & 0.55 & 0.97 \\
\projabbr-USP, U=8
& 1.02 & 0.98 & 0.93 & 0.92 & 1.21 & 2.22 & 1.43 & 2.26 \\
Ring-USP, U=16
& 1.66 & 1.92 & 2.29 & 2.50 & 0.69 & 0.97 & 0.69 & 0.97 \\
\projabbr-USP, U=16
& 1.05 & 0.96 & 0.94 & 0.91 & 0.96 & 1.67 & 1.18 & 2.00 \\
Ring-USP, U=32
& 1.15 & 1.15 & 0.93 & 1.17 & 0.69 & 0.97 & 0.69 & 0.97 \\
\projabbr-USP, U=32
& 1.07 & 0.95 & 0.95 & 0.90 & 0.78 & 1.75 & 0.90 & 1.48 \\
\hline
\end{tabular}
}
\end{table}

\subsection{Execution Time Breakdown Analysis}

\begin{figure}[!htbp]
    \centering
     \includegraphics[width=1.0\linewidth]{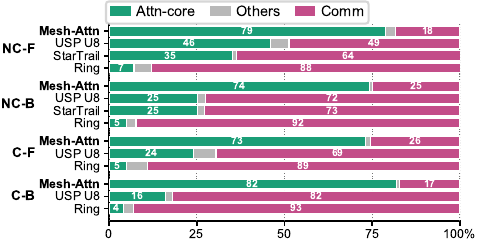}
      \vspace{-6mm}
  \caption{\rev{Execution-time breakdown (256 GPUs, 512K sequence length)}}
  \vspace{-3mm}
  \label{fig:time_breakdown}
\end{figure}

\rev{
Figure~\ref{fig:time_breakdown}
presents runtime breakdown analysis,
including core attention computation (Attn-core), auxiliary non-attention computation (Others), and communication overhead on the critical path (Comm). Across the evaluated settings, Comm accounts for only 21.5\% of \proj's runtime on average, compared with 90.6\%, 68.1\%, and 68.3\% for Ring-Attention, USP, and StarTrail, respectively. These results show that \proj's performance advantage stems primarily from reducing exposed communication overhead, while the baselines remain communication-bound.
}

\subsection{Communication Analysis}\label{sec:communication_analysis}
\begin{table}[t]
\centering
\small
\caption{
\rev{Communication analysis (256 GPUs, 512K sequence length).
(unit for volume: GiB)}
}
\label{tab:communication-volume}

\scalebox{0.86}{
\begin{tabular}{@{}clrrr@{}}
\hline
Mode & Method & Norm.Comm & Total.V & Inter.V \\
\hline

\multirow{4}{*}{\textbf{NC-F}}
& \projabbr  & 1.0   & 1.9  & 0.6 \\
& Ring       & 95.8  & 15.9 & 2.0 \\
& USP        & 6.9   & 2.0  & 1.9 \\
& StarTrail  & 11.8  & 2.9  & 2.0 \\
\hline

\multirow{4}{*}{\textbf{NC-B}}
& \projabbr  & 1.0  & 3.8  & 1.3 \\
& Ring       & 60.7 & 31.9 & 4.0 \\
& USP        & 7.6  & 4.0  & 3.9 \\
& StarTrail  & 8.0  & 5.8  & 2.3 \\
\hline

\multirow{3}{*}{\textbf{C-F}}
& \projabbr  & 1.0   & 2.4  & 0.6 \\
& Ring       & 112.9 & 15.9 & 2.0 \\
& USP        & 11.6  & 2.0  & 1.9 \\
\hline

\multirow{3}{*}{\textbf{C-B}}
& \projabbr  & 1.0   & 3.8  & 1.3 \\
& Ring       & 163.5 & 31.9 & 4.0 \\
& USP        & 20.8  & 4.0  & 3.9 \\
\hline
\end{tabular}}
\end{table}

\rev{
Table~\ref{tab:communication-volume} presents a detailed communication analysis. Norm.Comm denotes the critical-path communication time normalized to \proj, while Total.V/Inter.V report the per-GPU total and inter-node communication volumes. 
We estimate critical-path communication time 
by subtracting the 
profiled
compute time from the end-to-end
runtime.
Across the evaluated settings, \proj reduces the critical-path communication time by $101.8\times$, $10.6\times$, and $9.7\times$ on average compared with Ring-Attention, USP, and StarTrail. The corresponding reductions in total communication volume are $7.98\times$, $1.02\times$, and $1.52\times$, while the inter-node volume is reduced by $3.17\times$, $3.10\times$, and $2.37\times$.}

\rev{
It is worth noting that \proj reduces critical-path communication time by much more than it reduces communication volume
since \proj's overlapping schedule
only adopts efficient coarse-grained library-optimized collective primitives
while Ring-Attention, USP, and StarTrail
mostly relies on fine-grained P2P 
to hide communication cost.
For NC-F, \proj's final schedule contains eight all-gathers,
four reduce-scatters, and eight lightweight LSE all-reduces. In contrast, Ring-Attention comprises 255 sequential P2P exchange rounds, repeatedly incurring kernel-launch and synchronization overheads that reduce communication efficiency and make communication harder to hide. Worse still, cross-node P2P traffic is limited to the NIC attached to each endpoint GPU. Ring-Attention therefore uses only the two NICs at its inter-node ring boundaries (one for sending and the other for receiving cross-node traffic), whereas USP and \proj can exploit all eight NICs through independent Ring lanes and collectives, respectively. Together with Ring-Attention's fine-grained P2P kernels, this explains why Ring and USP have similar inter-node communication volumes but very different critical-path communication times.
}

\subsection{Evaluation with GQA}

Besides standard multi-head attention (MHA),
we also evaluate the performance of \proj 
for grouped-query attention (GQA)~\cite{ainslie2023gqatraininggeneralizedmultiquery} (Table~\ref{tab:gqa-runtime}),
an attention variant that slightly trades model expressiveness for 
better efficiency.
GQA reduces the number of KV heads by a factor of $g$ by allowing each group of $g$ consecutive query heads to share one KV head.

\begin{wraptable}[13]{r}{0.40\columnwidth}
\vspace{-0mm}
\centering
\small
\setlength{\tabcolsep}{1pt}
\caption{\rev{GQA runtime (NC-F, 256 GPUs, 512K, normalized to `\projabbr, GQA=1').}}
\label{tab:gqa-runtime}
\scalebox{1.0}{
\begin{tabular}{@{}lrrrr@{}}
\hline
GQA ($g$) & 1 & 2 & 4 & 8 \\
\hline
\projabbr     & \textbf{1.0} & \textbf{1.0} & \textbf{0.9} & \textbf{0.9} \\
Ring      & 20.1 & 10.2 & 7.8 & 7.6 \\
USP       & 2.7  & 1.7  & 1.4 & 1.3 \\
StarTrail & 3.5  & 2.2  & 1.6 & 1.3 \\
\hline
\end{tabular}}
\end{wraptable}

In practice, $g$ cannot be arbitrarily large (typically at most 8 or 16), as excessive sharing may degrade model quality. As $g$ increases, the KV tensors become smaller, reducing KV communication and alleviating the communication bottlenecks of baselines.
\proj consistently outperforms all baselines across the evaluated GQA ratios. Its advantage narrows at larger ratios (e.g., $g=4$ or $8$) because GQA reduces the baselines' communication overhead. However, GQA provides only a constant-factor reduction in communication volume (e.g., a factor of $g$ for Ring-Attention) and does not change their asymptotic communication complexity. We therefore expect \proj's advantage to widen again when scaling to larger GPU counts, such as 512 or 1024 GPUs.

\subsection{\proj Ablation Studies}

\noindent\textbf{Analyzing \proj factorization.}  We further study how different factorizations affect \proj by varying the Q group size on 128 GPUs with a 256K sequence. 
\begin{wrapfigure}[15]{r}{0.45\columnwidth}
    \centering
     \vspace{-0mm}
    \includegraphics[width=\linewidth]{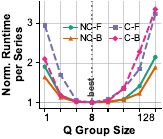}
    \caption{\rev{Sensitivity tests on Q group sizes. Each curve is independently normalized to its minimum runtime.}}
    \vspace{0mm}
    \label{fig:group_size_q_ablation}
\end{wrapfigure}
For each Q group size, we sweep the remaining configuration parameters and report the fastest configuration in Figure~\ref{fig:group_size_q_ablation}. Across all four phases, a Q group size of 8 achieves the lowest runtime, while extreme values such as 1 or 128 perform substantially worse.

\noindent\textbf{Analyzing \proj optimizations.} 
\rev{
Table~\ref{tab:feature-ladder} analyzes the contributions of three \proj optimizations: KV Partition Rotation (KVR), Topology-Aware Mapping (TAM), and Greedy Overlapping Optimization (GOO). We start with the best configuration identified for 128 GPUs and a 256K sequence (non-causal), disable all three optimizations to obtain the Basic implementation, and then restore them incrementally. Each optimization improves performance by reducing or hiding communication overhead. KVR is particularly effective because it makes communication symmetric and balanced across GPUs, allowing highly-optimized collectives (e.g., all-gather) to replace less efficient P2P transfers.
}

\begin{table}[t]
\centering
\footnotesize
\caption{
\rev{
\proj optimization ablations (128 GPUs, 256K sequence length, non-causal).
The runtime and communication time on critical path are normalized.}
}
\label{tab:feature-ladder}

{\setlength{\tabcolsep}{2pt}
\begin{tabular}{@{}clccc@{}}
\hline
Pass & Version & Norm. runtime & Norm. Comm. & Speedup \\
\hline

\multirow{4}{*}{\textbf{Forward}}
& Basic             & 1.00 & 1.00 & 1.00x \\
& w/ KVR            & 0.28 & 0.16 & 3.60x \\
& w/ KVR, TAM       & 0.22 & 0.09 & 4.60x \\
& w/ KVR, TAM, GOO  & \textbf{0.19} & \textbf{0.04} & \textbf{5.42x} \\
\hline

\multirow{4}{*}{\textbf{Backward}}
& Basic             & 1.00 & 1.00 & 1.00x \\
& w/ KVR            & 0.44 & 0.26 & 2.26x \\
& w/ KVR, TAM       & 0.36 & 0.15 & 2.78x \\
& w/ KVR, TAM, GOO  & \textbf{0.33} & \textbf{0.06} & \textbf{3.03x} \\
\hline
\end{tabular}
}
\end{table}

\noindent\textbf{Analyzing overlapping methods.} \rev{
We compare GOO with the two strawman overlap strategies described in
Section~\ref{sec:overlapping} using 128 GPUs, a 256K sequence, and non-causal
attention. GOO outperforms AGQ-RingKV by $1.82\times$ and $1.55\times$ in
forward and backward, respectively, and outperforms AGKV-RingQ/O by
$1.78\times$ and $1.19\times$. This advantage comes from coordinating overlap
across both the Q and KV dimensions, whereas each strawman pipelines only one
dimension.
}

%% file: conclusion.tex
\section{Conclusion}

This paper presents \proj, a distributed attention algorithm that balances Q and KV locality by assigning each GPU a two-dimensional tile of the Q-KV interaction matrix rather than a one-dimensional row or column, thereby reducing communication. \proj introduces three key optimizations that regularize and balance communication, overlap communication with computation, and reduce traffic over slower inter-node links. 
Our theoretical analysis shows that \proj achieves asymptotically lower communication volume than Ring-Attention while retaining unrestricted sequence parallelism.
Extensive experiments convincingly confirm its performance and scalability advantages over state-of-the-art
methods.